\documentclass[12pt]{iopart}
\usepackage{iopams}

\usepackage{bm}
\usepackage{epsfig}

\begin{document}

\title{Antiferro--quadrupolar structures in UPd$_3$ inferred from x-ray resonant Bragg diffraction}

\author{Javier Fern\'{a}ndez-Rodr\'{\i}guez}

\address{European Synchrotron Radiation Facility, BP 220, 38043  Grenoble Cedex, France}

\author{Stephen W. Lovesey}

\address{ISIS Facility, Rutherford Appleton Laboratory, Oxfordshire OX11 0QX, United Kingdom} 

\address{Diamond Light Source Ltd. Oxfordshire OX11 0DE, United Kingdom}

\author{Jes\'{u}s A. Blanco}
\address{Departamento de F\'{\i}sica, Universidad de Oviedo,
E-33007 Oviedo, Spain}

\date{\today}

\begin{abstract}

A systematic analysis of resonant x-ray Bragg diffraction data for UPd$_3$,
with signal enhancement at the U M$_{IV}$ edge, including possible structural
phase-transitions leads to a new determination of the space groups of the
material in the phases between $T_0 = 7.8$~K and $T_{+1}$ = 6.9 K as P222$_1$
and between $T_{-1}$ = 6.7 K and $T_{2}$ = 4.4 K as space group P2$_1$. In
addition, the quadrupolar order-parameters, $\langle Q_{ab}\rangle$, inferred from
diffraction data in the phase between $T_{-1}$ and $T_{2}$, are $\langle Q_{xz}\rangle$ and
$\langle Q_{yz}\rangle$ at the (103) Bragg reflection, and $\langle Q_{xy}\rangle$ at the (104)
reflection.

\end{abstract}

\maketitle

\section{Introduction}

Phase transitions in materials are often driven by a cooperative
action among degrees of freedom of the electrons. The
paramagnetic-magnetic transition, for example, may arise from
interactions between magnetic dipole moments. The potential importance
of multipolar, e.g., quadrupolar, interactions in f-electrons system
has long been recognized \cite{Morin90,
Santini2009,Kuramoto2009}. Experimental techniques utilizing resonant
x-ray Bragg diffraction (RXS), which have evolved in the past decade,
are now the preferred techniques for exposing the interplay of charge,
spin and orbital electron degrees of freedom in complex materials.  In
short, RXS reveals multipoles with a directness of purpose not
available with any other technique in the science of materials. By
measuring intensities of weak, space-group forbidden reflections,
families of multipoles can be detected, including, magnetic charge (or
magnetic monopole)~\cite{Lovesey2009}, dipole~\cite{Fernandez2005},
anapole~\cite{Lovesey2007b,Fernandez2009},
quadrupole~\cite{Wilkins2006}, octupole~\cite{Paixao2002,Lovesey2000}
and hexadecapole~\cite{Tanaka2004,Fernandez2008}.  Use of resonant
x-ray Bragg diffraction to detect some of these strange multipoles
hosted by complex materials is quite analogous to atomic parity
violation experiments to measure the nuclear anapole
\cite{Tsigutkin2009} and it is the method by which to verify strange
multipoles predicted in simulations \cite{Cricchio2009}.

A great deal of attention has focused on transition-metal
materials especially manganites, in which it has been shown that orbital
order is the key to understanding a raft of relevant physical properties
\cite{Millis98,Ahn92}. The situation of rare-earth (4f) or actinide (5f) ions is
different because the coupling between the spin and orbital angular
momentum is relatively strong. High-order multipolar interaction may
eventually manifest themselves in other subtle effects, such as lifting the
degeneracy between single or multi-k structure, or independently of any
magnetic order, the interactions between the 4f shells drive orderings in
which their electronic density no longer respect the initial symmetries. In
the latter cases, the primary order parameter is not of magnetic nature,
but of orbital one. As the additional 4f asphericity is usually well
described by the emergence of a quadrupolar arrangement, this modification
of the system is called quadrupolar order. The case of multipolar ordering,
although its possibility has long been recognized and theoretically
investigated, remains a rather esoteric issue mainly because of the limited
number of systems that unambiguously display the behaviour \cite{Morin90}. The
situation has improved recently because novel experimental findings have
been reported for several more rare-earth and actinides compounds
\cite{MuldersPRB,Fernandez2009JPCM,Kuramoto2009}.

The purpose of this article is to interpret resonant x-ray Bragg
diffraction by UPd$_3$ \cite{McMorrow2001,Walker2006}, in particular
the dependence of the intensities of the different super-lattice peaks
on sample temperature and sample orientation (azimuthal angle) in
terms of the possible lowerings of space-group symmetry and their
associated quadrupolar structures. Russell-Saunders coupling scheme
yields $S = 1$, $L = 5$, $J = L - S = 4$. The large orbital moment
gives rise to a strong coupling to the lattice and hence it is not
surprising that this is a system in which quadrupolar effects are
dominant.  Because the U -U distance, which is around $4.11$~\AA, is
larger than that of the Hill limit (3.6 \AA ), the system (UPd$_3$)
behaves like a localised 4f-system. The low temperature specific heat
of UPd$_3$ reveals a contribution from low-lying crystal-field (CF)
levels, which gives rise to sharp phase transitions. The order would
then results from a delicate balance between rather strong magnetic
interactions and CF interactions.  The possibility that quadrupolar
order occurs in UPd$_3$ was first raised in \cite{Andres78} and
reiterated since then.

McEwen and co-workers \cite{McEwen2003JPCM} re-analysed available
experimental data for UPd$_3$, and produced both a new energy level scheme,
with wave-functions, for the high-temperature phase, and a model for the
progression of the phase transitions.

\section{Calculation of the RXS structure factor}

Resonant x-ray Bragg diffraction experiments
\cite{McMorrow2001,Walker2006} that we discuss were carried with the
x-ray energy tuned to the M$_{IV}$ edge of U at 3.728 keV. At this
edge, the resonant scattering cross-section is dominated by electric
dipole (virtual) transitions connecting 3d$_{3/2}$ and 5f$_{5/2}$
electron states. The observation of satellite peaks at the M$_{IV}$
edge of U establishes both that long-range order of the 5f electrons
occurs in UPd$_3$ and that these experiments probe ordering of the
quadrupolar moments in the 5f shell.

For Bragg diffraction enhanced by an electric dipole event (E1)
the unit-cell structure factor, F, can be written as a product of a
quantity $\mathbf{X}$, which describes the condition of the x-rays, and a quantity $\bm{\Psi}$
that is related to the valence electrons \cite{Lovesey2005}

\begin{equation}
F = \sum_K \mathbf{X}^K \cdot \Psi^K = \sum_{K,Q} (-1)^Q X^K_{-Q} \Psi^K_Q.
\end{equation}

Here, $K = 0, 1, 2$..., is the rank of a spherical component and the
projection Q takes values within the range $-K < Q < K$. $X^K_Q$ is
described in terms of coordinates fixed by the chosen geometry of the
experiment, while $\Psi^K_Q$ is written with respect to axes in the crystal.
$\Psi^K_Q$,  a structure factor of electrons in a unit cell, is a sum of
multipoles $\langle T^{K}_{Q}\rangle_{\mathbf{d}}$, at sites
$\mathbf{d}$ in the unit cell multiplied by the usual
spatial phase factors,

\begin{equation}
\Psi^K_Q = \sum_{\mathbf{d}} \exp(i \mathbf{k\cdot d}) \langle T^{K}_{Q}\rangle_{\mathbf{d}}     .
\end{equation}

Point-group symmetry places restrictions on the allowed components of
$\langle T^{K}_{Q}\rangle$. Measured intensities reveal information on
anisotropy in time-even, charge-like (K = even integer) and time-odd,
magnetic (K = odd integer) distributions. For scattering enhanced by
an E1 event one can observe multipoles with rank K = 0 (monopole or
charge), K = 1 (dipole) and K = 2 (quadrupoles).

\section{Properties of UPd$_3$}

At room temperature, the material crystallizes in the double hexagonal
close-packed structure (dhcp) P6$_{3}$/mmc with U ions at sites with
locally quasi-cubic (2a Wyckoff positions) and hexagonal (2d Wyckoff
positions) symmetry. The palladium ions are at 6g and 6h positions.
Benefiting from the selective character of RXS, Pd ions play no part
and the observed RXS intensity is coming from quasi-cubic uranium
ions, given the fact that ordered quadrupole moments are predominantly
at the quasi-cubic sites~\cite{McEwenJMMM98, McMorrow2001}. Resonant x-ray
diffraction measurements~\cite{McMorrow2001,Walker2006} were done at
reflections (103) and (104) in ortho-hexagonal axes with lattice
constants $a= \sqrt{3} a_{hex}$, $b =a_{hex}$, and $c= c_{hex}$.  In
order to align crystal coordinates with the x-ray coordinates defined
in \cite{Lovesey2005}, it is necessary to perform a rotation of
coordinates defined by the Euler angles $(\pi , \beta_{10l}, \pi )$ with
$\beta_{103} = 108.0$~degrees and $\beta_{104} =
103.7$~degrees. Measurements in \cite{McMorrow2001} were only done at
a fixed azimuthal angle $\psi$= 90 degrees.

Bulk measurements (heat capacity, electrical resistivity, ultra-sounds
measurements, magnetic susceptibility, among others) together with
x-ray and neutron scattering experiments~\cite{McEwenJMMM98} have
revealed up to four quadrupolar phase-transitions: $T_0= 7.8$~K,
$T_{+1} = 6.9$~K, $T_{-1} = 6.7$~K and $T_{2} = 4.4$~K, being the
transitions at $T_0$ and $T_{+1}$ second order and the transitions at
$T_{-1}$ and $T_2$ first order.  In the latter phase, the
antiferro-quadrupolar (AFQ) transition is accompanied by the
appearance of a small antiferromagnetic dipolar moment.

\section{Minimal symmetry model structure-factor}

We start by deriving the consequences of a minimal symmetry model
structure factor similar to the one used in refs. \cite{Lovesey2007a}
and \cite{Scagnoli2009}, and investigate the possibility of a 2-fold
axis of rotation symmetry about the b crystallographic axis. The
general structure factor in atomic axes $(\xi,\eta,\zeta)$ with zeta
parallel to the crystal b-axis, can be written as

\begin{equation}
\Psi^K_Q(\xi,\eta,\zeta) = B^K_Q +D^K_Q      
\end{equation}

with $B^K_{-Q} = B^K_Q$ and $D^K_{-Q} = -D^K_Q$. Atomic axes $(\xi,\eta,\zeta)$ are
related to crystal axes (a,b,c) by a rotation of $\pi/2$ about the a-axis. If
we calculate the resonant structure factor for rank 2 tensors (quadrupoles)
in reflections of the kind $(10l)$ at azimuthal angle of 90 degrees, we
obtain,
\begin{eqnarray}
F_{\sigma'\sigma} &=& \sqrt{\frac{3}{2}} B^2_0 \nonumber \\
F_{\pi'\sigma} &=& ReD^2_{1} cos (\beta+\theta) + ImB^2_{1} sin (\beta+\theta) \nonumber \\
F_{\pi'\pi} &=& ReB^2_{2} (2cos^2\beta -1) - \frac{1}{\sqrt{6}}B^2_0 \cos 2\theta 
 + ImD^2_{2} \sin 2\beta \nonumber \\
F_{\sigma'\pi} &=& ReD^2_{1} \cos (\beta-\theta) + ImB^2_{1} \sin
(\beta-\theta)  .
\end{eqnarray}
Assuming a point symmetry including a 2-fold axis about b-axis, $|Q|$ is an
even integer. Thus $B^2_{1}$ and $D^2_{1}$ would be forbidden leading to zero
intensity in the rotated channels $\pi'\sigma$ and $\sigma'\pi$ at $\psi$= 90 degrees.
This is compatible with the RXS measurements done at 7.1 K
\cite{Walker2006,McMorrow2001}. In the measurements done at T= 5.2 K, the
appearance of intensity in the rotated channel at the azimuthal angle $\psi$ = 90 degrees implies
that the 2-fold axis point symmetry about b is no longer present.

\section{RXS measurements at T = 7.1 K}

There is evidence that the material undergoes a structural
transition to an orthorhombic space group \cite{Zochowski94,Lingg1999}. Among the
orthorhombic subgroups of P6$_{3}$/mmc compatible with a 2-fold axis of point
symmetry along b for uranium ions occupying quasi-cubic sites, we find that
P222$_1$ reproduces the azimuthal measurements published in
ref.~\cite{Walker2006}.  In P222$_1$ the $(10l)$ reflections would became space group
allowed, in agreement with the weak non-resonant intensity in the unrotated
channel reported in \cite{McMorrow2001}. The quasi-cubic uranium ions would be
at positions 2c and 2d of P222$_1$. Uranium ions with $z = 1/4$ and $z = 3/4$
are related by a two-fold axis about c.  The quadrupolar structure factors
for the 2c and 2d sites for the reflection $(10l)$ and arbitrary Miller index
$l$ are,
\begin{eqnarray}
\Psi^2_Q  (10l)_{(2c)}&=& \exp(i \pi/2 l) (1+(-1)^{l +Q} ) \rangle
T^{2}_{Q}\rangle_{(2c)} \nonumber\\
\Psi^2_Q  (10l)_{(2d)}&=& -\exp(i \pi/2 l) (1+(-1)^{l+Q} ) \langle
T^{2}_{Q}\rangle_{(2d)}   .
\end{eqnarray}
In these structure factors, we use a different set of axes for the
multipoles than those used in previous section, for the axes of
$\langle T^{2}_{Q}\rangle$ are now aligned with crystal axes a, b, c.
At the reflection $(10l)$ what it is measured is the difference
between the quadrupolar moments in the $(0,0,0)$ and
$(\frac{1}{2},\frac{1}{2},0)$ orthorhombic positions, namely, 
$\langle T^{2}_{Q}\rangle_{(0,0,0)} - \langle T^{2}_{Q}\rangle_{(\frac{1}{2},\frac{1}{2},0)}$.  As they occupy
inequivalent Wyckoff positions (2c and 2d) quadrupoles at positions
$(0,0,0)$ and $(\frac{1}{2},\frac{1}{2},0)$ are not related by the
space-group.  A quadrupolar ordering in which $\langle
T^{2}_{Q}\rangle_{(0,0,0)} = \langle
T^{2}_{Q}\rangle_{(\frac{1}{2},\frac{1}{2},0)}$ would not be observed at
the $(10l)$ reflection.

The factor $(1+(-1)^{l +Q} )$ in (5) implies that multipoles $\langle
T^{2}_{Q}\rangle$ with $|Q | = 1$ are observed in the $(103)$
reflection, while $(104)$ would give access to multipoles with even $|Q
|$ ( $|Q |=0,2$).  In addition, point symmetry (with a 2-fold axis about b)
forces $\langle T^{2}_{Q}\rangle$ to be purely real, and this means
that in the $(103)$ reflection Re$\langle T^{2}_{1}\rangle \equiv \langle
Q_{zx}\rangle$ is observed while at the $(104)$ reflection Re$\langle
T^{2}_{2}\rangle \equiv \langle Q_{x^2-y^2}\rangle$ and $<T^2_0> \equiv
\langle Q_{3z^2-r^2}\rangle$ are observed.

For reflections of the kind $(10l)$ with arbitrary $l$ the total
structure factor will be proportional to the difference of multipoles
between the two crystal Wyckoff positions, $\langle
T^{K}_{Q}\rangle_{(2c)} - \langle T^{K}_{Q}\rangle_{(2d)}$, i.e.,
to the difference of the mean values of the multipoles between the sites
(0,0,0) and $(\frac{1}{2},\frac{1}{2},0)$ in orthorhombic coordinates.

The absence of measured resonant intensity at (104) at this
temperature \cite{McMorrow2001} implies that $\langle Q_{3z^2-r^2}\rangle$ and $\langle Q_{xz}\rangle$ have no
antiferro-ordering. The intensity of (103) reflection is proportional to
the difference of $\langle Q_{zx}\rangle$ between the  2c and 2d crystal positions,
$\langle Q_{zx}\rangle_{(2c)}- \langle Q_{zx}\rangle_{(2d)}$ and is given by

\begin{eqnarray}
F_{\sigma'\sigma(103)}&=&  -2 \langle Q_{zx} \rangle  \sin\beta_0
\cos\alpha_0 \\
F_{\pi'\sigma(103)}&=& -\langle Q_{zx} \rangle  
\big\lbrack  
\sin (\alpha_0+\gamma_0-\theta)(1+\cos\beta_0)(2\cos\beta_0-1))+
\nonumber \\
&&\sin (\alpha_0-\gamma_0+\theta)(1-\cos\beta_0)(2\cos\beta_0+1))
\big\rbrack 
\end{eqnarray}
The angles $\alpha_0$, $\beta_0$, $\gamma_0$ result from the composition the
azimuthal rotation and the rotation that aligns the Bragg wavevector with
the x-ray coordinates, and are given by
\begin{eqnarray}
\alpha_0 &=& \mathrm{arccot}\left(-\cot\psi\sin \beta \right) \nonumber \\
\beta_0  &=& \arccos\left(\cos\psi \cos\beta \right)         \nonumber   \\
\gamma_0 &=& \mathrm{arccot}\left(-\cot\beta\sin \psi \right)
\end{eqnarray}
with the angle $\beta$ for the (103) and (104) reflections being:
$\beta_{103} = 108.0$~degrees, $\beta_{104} = 103.7$~degrees.

The azimuthal dependence of the (103) reflection together with
experimental data \cite{Walker2006} is depicted in Fig. 1. We have
allowed contribution from a non-resonant (Thomson) term in the $\sigma'\sigma$
channel. The quadrupolar structure depicted in fig. 4 of
ref. \cite{Walker2006} corresponds to the antiferro-quadrupolar
ordering of $\langle Q_{xz}\rangle$ observed, i.e.  $\langle
Q_{xz}\rangle_{(2c)}=-\langle Q_{xz}\rangle_{(2d)}$. A possible
contribution of $\langle Q_{x^2-y^2}\rangle$ that would modify the
azimuthal dependence of the (103) reflection was proposed in
\cite{McEwen2007JMMM}. However, in space group P222$_1$ contribution of
$\langle Q_{x^2-y^2}\rangle$ to the (103) reflection is not allowed, as
it would have in-phase stacking along the c-axis, i.e. when making the
translation $(0,0,\frac{1}{2})$ along z, the value of $\langle
Q_{x^2-y^2}\rangle$ would be the same (this is different to the
structure depicted in \cite{McMorrow2001} which has antiphase stacking
of $\langle Q_{x^2-y^2}\rangle$ along c-axis). Still, in the structure
of space group P222$_1$ there would the possibility of ordering of
$\langle Q_{x^2-y^2}\rangle$ with $\langle Q_{x^2-y^2}\rangle_{(000)}$ =
$\langle Q_{x^2-y^2}\rangle_{(\frac{1}{2},\frac{1}{2},0)}$, which would not
be observable in $(10l)$ reflections.

\begin{figure}
\begin{center}
\includegraphics[width=4in]{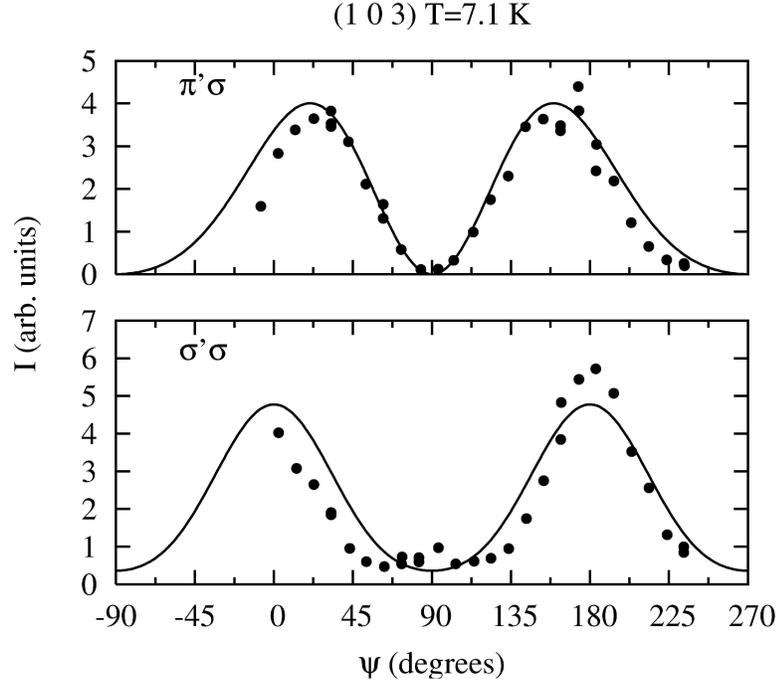}
\end{center}
\caption{\label{Fig1} Azimuthal variation of the intensity in the (103) reflection
measured at T= 7.1 K. Experimental points are taken from 
ref.~\cite{Walker2006}. The
continuous line shows the fit of the data to eqs. (6) and (7).} 
\end{figure}

\section{RXS measurements at T = 5.2 K}

To accommodate the observed change to the shape of the (103) azimuthal
curve, and the appearance of intensity in $\pi'\sigma$ at 90 degrees
in the azimuthal curves of both (103) and (104) reflections (see
Figs. 2 and 3), it is necessary to reduce the symmetry experienced by
the quasi-cubic uranium ions (two-fold axis about b). Possible
subgroups of P222$_1$ that do no break the translational symmetry are
P2$_1$, P2, P1.  The possibility of a lowering of symmetry to a
monoclinic space group has already been suggested in
ref.~\cite{Lingg1999}.  The minimum reduction of symmetry to justify
these changes would be to consider the monoclinic space group P2$_1$.
In this space group, ions at different positions in $z$ within the
unit cell continue to be related by a two-fold axis about c. The
quasi-cubic uranium ions in this group would occupy two different sets
of 2a Wyckoff positions. The algebraic form of unit-cell structure
factors is unchanged, but now there is no point symmetry and $\langle
T^{K}_{Q}\rangle$ is no longer purely real. In consequence, there are
additional contributions containing $\langle Q_{yz}\rangle$ at (103)
and $\langle Q_{xy}\rangle$ at (104). The azimuthal dependence of the
(103) reflection is then:
\begin{eqnarray}
F_{\sigma'\sigma(103)}&=&-2\sin\beta_0(  \langle Q_{zx} \rangle \cos\alpha_0+\langle Q_{yz} \rangle\sin\alpha_0)
\\
F_{\pi'\sigma(103)}&=&   -\langle Q_{zx} \rangle
\big\lbrack  
\sin (\alpha_0+\gamma_0-\theta)(1+\cos\beta_0)(2\cos\beta_0-1))+
\nonumber \\
&&\sin (\alpha_0-\gamma_0+\theta)(1-\cos\beta_0)(2\cos\beta_0+1))
\big\rbrack +
\nonumber \\
&&\langle Q_{yz} \rangle 
\big\lbrack  
\cos(\alpha_0+\gamma_0-\theta)(1+\cos\beta_0)(2\cos\beta_0-1))-
\nonumber \\
&&\cos(\alpha_0-\gamma_0+\theta)(1-\cos\beta_0)(2\cos\beta_0+1))
\big\rbrack        ,
\end{eqnarray}
and for the (104) reflection,
\begin{eqnarray}
F_{\sigma'\sigma(104)}&=& \langle Q_{zz} \rangle
(3\cos^2 \beta_0-1) \nonumber \\
&&+2\sin^2\beta_0 ( \frac{1}{2}(\langle Q_{xx}-Q_{yy} \rangle ) \cos 2\alpha_0+\langle Q_{xy} \rangle\sin 2\alpha_0) \\
F_{\pi'\sigma(104)}&=&-\frac{3}{2} \langle Q_{zz} \rangle
\sin 2\beta_0 \sin (\gamma_0-\theta) +\nonumber \\
&&\sin\beta_0(1+\cos\beta_0)
\nonumber \\
&&\Big\lbrace
\frac{1}{2}(\langle Q_{xx}-Q_{yy} \rangle ) \rangle \lbrack (1+\cos\beta_0)\sin(2\alpha_0+\gamma_0-\theta)+\nonumber \\
&& (1-\cos\beta_0)\sin(2\alpha_0-\gamma_0+\theta)\rbrack - \nonumber \\
&&\langle Q_{xy} \rangle \lbrack
(1+\cos\beta_0)\cos(2\alpha_0+\gamma_0-\theta)+ \nonumber \\
&& (1-\cos\beta_0)\cos(2\alpha_0-\gamma_0+\theta)\rbrack                    
\Big\rbrace             .
\end{eqnarray}
Fig. 2 shows the fitting of the available data for (103)
\cite{Walker2006} with $\langle Q_{zx}\rangle$ and $\langle
Q_{yz}\rangle$ in (9) and (10) allowed to be different from zero,
together with the other quadrupoles. The value of $\langle
Q_{zx}\rangle$/$\langle Q_{yz}\rangle$ from the fitting is: $\langle
Q_{zx}\rangle$/$\langle Q_{yz}\rangle = - 4.72 \pm 0.12$. Measurements
done with $\pi$ incident polarization \cite{McMorrow2001} show that at
T=5.2 K, at $\psi$=90 degrees, in (103) $\pi'\pi$ is about one order of
magnitude more intense than $\sigma'\pi$ (with our derived parameters we
obtain $I_{\pi'\pi}/I_{\sigma'\pi} = 19.4$ at $\psi =  90$ deg. and
$I_{\pi'\pi}/I_{\sigma'\pi} = 8.3$ at $\psi = -90$~deg.).  This differs from the
fit done in \cite{Walker2006}, which attributed the variation in the
azimuthal dependence in (103) at 5.2 K solely to the contributions of
$\langle Q_{xy}\rangle$ and $\langle Q_{x^2-y^2}\rangle$.  In our
proposed structure these two quadrupoles have in-phase stacking along
c-axis and do not contribute to (103) reflection.

\begin{figure}
\begin{center}
\includegraphics[width=4in]{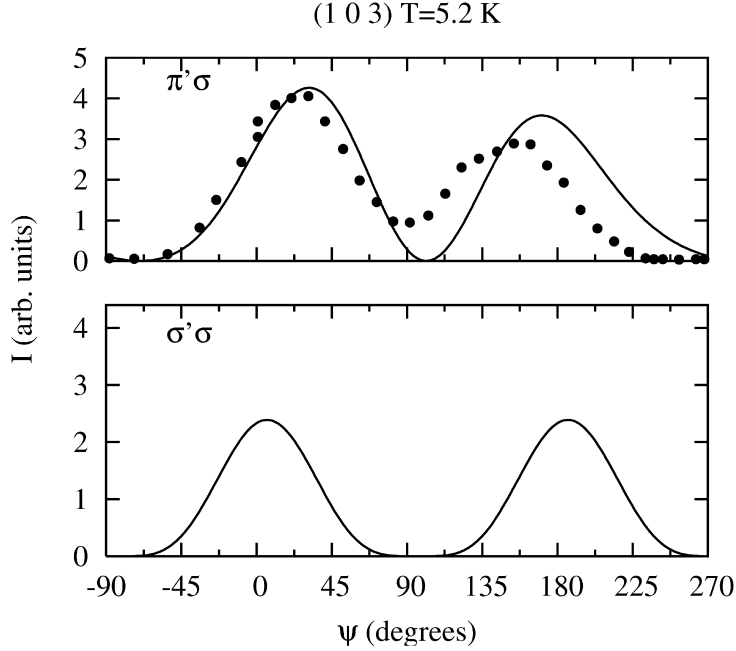}
\end{center}
\caption{\label{Fig2}Azimuthal variation of the intensity in the (103)
reflection measured at T= 5.2 K. Experimental points are taken from
ref.~\cite{Walker2006}. Continuous lines show the fit of the data to eqs. (9)-(10).}
\end{figure}

Available data \cite{Walker2006,Walker2008} for the (104) reflection
is shown in Figure 3 together with the azimuthal dependence
calculated from (11) and (12) with only $\langle Q_{xy}\rangle$
allowed to be different from zero and a non-resonant contribution
(Thomson) in $\sigma'\sigma$ that is independent of the azimuthal
angle. Inclusion of $\langle Q_{x^2-y^2}\rangle$ and $\langle
Q_{zz}\rangle$ does not lead to a significant improvement in the fit
to (104) data.  Moreover, in the case of the (104) reflection
measurements done with $\pi$ incident polarization at T = 5.2 K
\cite{McMorrow2001} show a very weak intensity in the $\pi'\pi$ channel
and both $\langle Q_{x^2-y^2}\rangle$ and $\langle Q_{zz}\rangle$ give a
non-zero intensity at $\psi$=90 degrees in $\pi'\pi$.  However, an
accidental cancellation could occur if $\langle Q_{xx}-Q_{yy}\rangle =
3\langle Q_{zz}\rangle$.

\begin{figure}
\begin{center}
\includegraphics[width=6in]{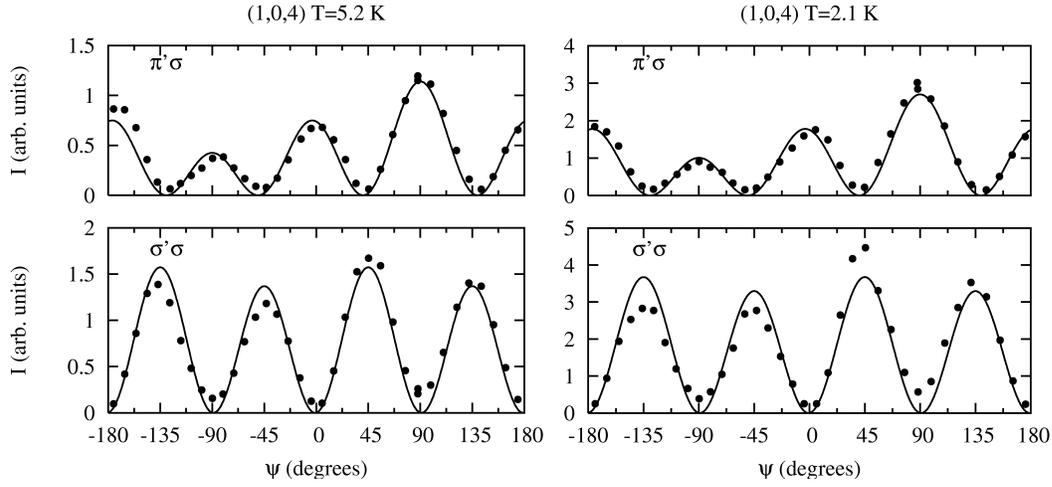}
\end{center}
\caption{\label{Fig3} Azimuthal variation of the intensity in the (104) reflection
measured at T= 5.2 K and 2.1 K. Experimental points are taken from
ref.~\cite{Walker2008}. Continuous lines show the fit of the data to eqs. (11) and
(12).}
\end{figure}

Figure 4 shows the antiferro-quadrupolar orderings of $\langle Q_{yz}\rangle$
and $\langle Q_{xy}\rangle$ derived from the proposed space-group structure, which are
compatible with data collected for the (103) and (104) reflections at T =
5.2 K and coexist with the ordering of $\langle Q_{zx}\rangle$ observed at 7.1 K
\cite{Walker2006}. The stacking along the c-axis of the $\langle Q_{yz}\rangle$ component of the
quadrupolar moment shown in Fig. 4-a changes sign when making a translation
of 1/2 of the z coordinate, which is different to the $\langle Q_{yz}\rangle$ stacking used
in refs. \cite{Walker2006} and \cite{Walker2008}.

\begin{figure}
\begin{center}
\includegraphics[width=4.5in]{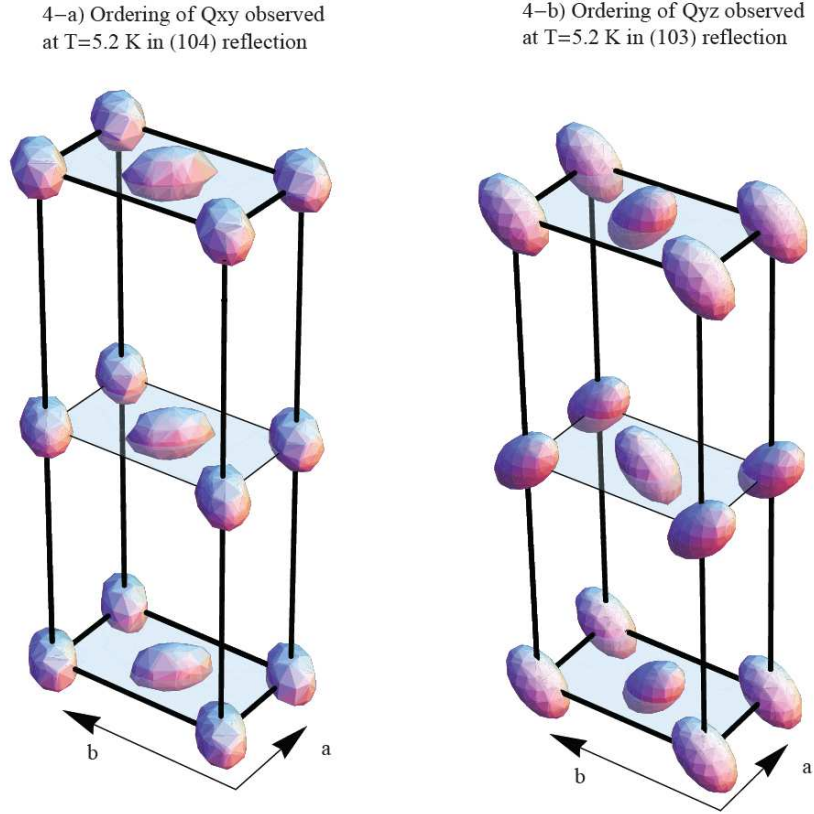}
\end{center}
\caption{\label{Fig4}Antiferroquadrupolar orderings of $\langle
Q_{xy}\rangle$ and $\langle Q_{yz}\rangle$ observed in the (104) and
(103) reflections at T = 5.2 K.}
\end{figure}

\section{Conclusions}

Resonant x-ray Bragg diffraction data are evidence of an
antiferro-quadrupolar ordering in UPd$_3$. Our analysis of available
data identifies the space group for the phase between $T_0 = 7.8$~K
and $T_{+1}$ = 6.9 K as P222$_1$.  Additionally, we can infer the
quadrupole components $\langle Q_{\alpha\beta}\rangle$
($\alpha,\beta=x,y,z$) and their stacking pattern contributing to
different space-group forbidden reflections. Intensity in the (103)
reflection is here assigned to the $\langle Q_{zx}\rangle$
quadrupole.  For the phase below $T_{-1}$ = 6.7 K, measurements are
compatible with the space group P2$_1$, which implies loss of the
2-fold axis of rotation symmetry in the point symmetry sites used by
the U ions.  However, we should be cautious with this assignment of
space group, given that the phase transition at $T_{-1}$ is
first-order and in that case the low temperature space-group is not
necessarily a subgroup of P222$_1$.  The intensity at the (103)
reflection would come mainly from $\langle Q_{zx}\rangle$ together
with a small $\langle Q_{yz}\rangle$ contribution. Intensity at (104)
is attributed to $\langle Q_{xy}\rangle$ with a possible small
contributions from $\langle Q_{x^2-y^2}\rangle$ and $\langle
Q_{zz}\rangle$. Within the derived structures for both phases of the
compound, no additional contributions from other quadrupolar
parameters are present in (103) and (104) reflections.

\section{Acknowledgements}

We are grateful to K. McEwen and D. McMorrow for useful discussions. One of
us, J.F.R., is grateful to Gobierno del Principado de Asturias for the
financial support of a Postdoctoral grant from Plan de Ciencia, Tecnologia
e Innovacion (PCTI) de Asturias 2006-2009. Financial support from
FEDER-MICCIN MAT2008-0654-C04-03 is also acknowledged.

\section{References}

\bibliography{UPd3_2009}
\bibliographystyle{unsrt}

\end{document}